\documentclass[preprint,12pt]{elsarticle}

\usepackage{amssymb}
\usepackage{amsmath}

\journal{Physics Letters A}

\begin{document}

\begin{frontmatter}



\title{Investigation of the Aharonov-Bohm and Aharonov-Casher Topological Phases for Quantum Entangled States in 2+1 Dimensions}


\author{H.O. Cildiroglu$^{*,a}$}

\affiliation{{Department of Physics Engineering},
            addressline={Ankara University}, 
            city={Ankara},
            postcode={06100}, 
            country={Turkey}}

\author{A.U. Yilmazer$^{a}$}

\begin{abstract}
Aharonov-Bohm (AB) and Aharonov-Casher (AC) effects are treated fully relativistically in 2+1 dimensions. The influences of the relevant geometric and topological phases on an entangled spin-1/2 system are studied. It is shown that for the AC case the correlation function of the Clauser-Horne-Shimony-Holt (CHSH) inequality for certain choices of the spin measurements depends on the AC phase explicitly. 
\end{abstract}


\end{frontmatter}


\section{Introduction}
\label{sec:sample1}
Geometric phases that occur in both classical and quantum systems adiabatically evolving through a closed cycle defined in the configuration space of its external parameters have been integrated into a unifying framework with the significant work of M. Berry \cite{1}. Special cases of geometric phases arise in the case of some topological singularities in the parameter space. The first and the most well-known example of this type is the physical scattering process of Aharonov-Bohm (AB) introduced in 1959 \cite{2}. The comprehensive work of Aharonov and Bohm involves examining the motions of electrons around the singular region created by a magnetic field due to an infinitely thin solenoid. A similar effect was discovered by Aharonov and Casher (AC-phase) in 1984 \cite{3}; here instead of a solenoidal magnetic field an electrical field due to an infinitely long and thin linear charge density is considered and also neutrons with non-zero magnetic dipole moments take the place of scattered electrons.

On the other hand, the entanglement which may appear in quantum systems having two or more subparts constitutes one of the deepest aspects of quantum mechanics. The state of two or more particle systems in quantum mechanics is generally expressed as linear superpositions of tensor product of discrete states. In this context, entanglement is connected to superpositions that cannot be expressed by a single tensor product, thus it can be seen as a new potential for the quantum states to exhibit correlations that cannot be calculated classically. The concept of the entanglement was used by Einstein, Podolsky and Rosen (EPR) in their article in 1935 to criticize the Copenhagen interpretation of quantum mechanics via the assigned predetermined values for the physical quantities \cite{4,5}. Since their gedanken experiment defies the locality and reality principles, this argument based on the EPR paradox has consequently initiated the road to the hidden variable theories. In his seminal work on the EPR problem, John Bell derived an inequality based on the local realistic hidden variables, known today as Bell inequality (BI), that is violated by quantum mechanics \cite{6}. This important development was followed by the study of Clauser, Horne, Shimony and Holt (CHSH) in 1969 which allowed an easy experimental test between local hidden variable models and quantum mechanics \cite{7}. The experimental violation of Bell-CHSH inequalities demonstrated that the nature itself might be fundamentally nonlocal and the quantum theory turned out to be the best description of the physical reality \cite{8,9}. 

In order to set these phenomena into a more systematic and unified framework and also to uncover the relationships between different quantum mechanical processes, many studies have been carried out [10-45]. \nocite{10,11,12,13,14,15,16,17,18,19,20,21,22,23,24,25,26,27,28,29,30,31,32,33,34,35,36,37,38,39,40,41,42,43,44,45} Bertlmann and Durstberger et. al. discussed the effects of the geometric Berry phase on the entangled spin-1/2 system in 2004 [25]\nocite{25}. In an entangled system, they have analyzed how the Berry phase affects the Bell angles and the maximum violations of BI. From this point of view quantum mechanical systems can further be investigated in such EPR type problems via studying the influence of the topological phases for the entangled states of two particles with spins. In this study a correlation relation in connection with the AC phase will be derived, providing a criterion which is maximally violated by the quantum mechanics; and it can be used in the future experiments to test the non-local properties of quantum mechanics in entangled states.

\section{AB Effect for Entangled Quantum Systems}
\label{sec:sample2}

The AB effect can simply be explained via the gauge potentials in the framework of non-relativistic quantum mechanics. The crucial idea is to restore the gauge invariance of non-relativistic quantum mechanics; hence the wavefunction must also acquire a phase related to the line integral of the gauge potential. However, it is beneficial to study the problem within the framework of relativistic quantum mechanics in order to investigate the effects of phase on entangled quantum systems. It would be appropriate to start from the relativistic Lagrangian which describes the dynamics of a single electron in the electromagnetic fields.
\begin{equation}
L=\bar{\psi }\lbrack i\gamma ^{\mu }\partial _{\mu 
}-m+e\gamma ^{\mu }A_{\mu }\rbrack \psi
\end{equation}

\noindent The equations of motion to be obtained are,
\begin{equation}\lbrack i\gamma ^{\mu }\partial _{\mu }-m+e\gamma^{\mu }A_{\mu }\rbrack \psi =0.\end{equation}

The experimental setup and the nature of the AB effect is surely three dimensional, however the translational invariance of the problem along the $z$ direction allows us to analyze the problem in 2+1 dimensions. When the problem confined on a plane, the phase is revealed by the motion of electrons in closed orbits formed around a polarized magnetic dipole since the solenoid can be thought of as a continues sequence of polarized magnetic dipoles placed on top of each other. In this context let us first consider the following free Dirac Hamiltonian,
\begin{equation}
    H_{D}=\alpha _{x}p_{x}+\alpha _{y}p_{y}+m\beta .
\end{equation}

As is well known in addition to the usual gamma matrix representation in 3+1 dimensional space-time the Clifford algebra $\{\gamma ^{\mu },\gamma ^{\nu } \}=2g ^{\mu\nu }$, has a particular two-dimensional representation in 2+1 dimensional space-time as given by:

\begin{equation}
\begin{matrix}
\alpha_{x}=\sigma_{x} & \\
\alpha_{y}=s\sigma_{y} & \\
\beta=\gamma_{0}=\sigma_{z} & \\
\end{matrix}
\end{equation}

\noindent Here $s=\pm 1$ for spin up and down polarizations and 
\begin{equation}
\begin{matrix}
\gamma _{1}=\beta \alpha _{x}=\sigma _{z}\sigma _{x}=i\sigma _{y}  \\
\gamma _{2}=\beta \alpha _{y}=s\sigma _{z}\sigma _{y}=-is\sigma _{x}.
\end{matrix}
\end{equation}

In the AB case, there is a presence of a magnetic field while the electric field is zero. Hence, four-vector-potential has the components $A_{0}=A_{z}=0$; also if the space and time elements of the covariant derivative in (2) are written separately and multiplied by $\gamma ^{0}$ one can get the effective AB Hamiltonian as
\begin{equation}\Delta H_{AB}=-e\boldsymbol{\alpha _{\perp}}\cdot\boldsymbol{A} \end{equation}

\noindent The corresponding eigenvalues, $\lambda _{AB_{1,2}}=\pm \thinspace eA$,
\noindent with the instantaneous eigenstates along the arbitrary $\hat{n}$ direction on the xy plane (see Fig. 1),
\begin{equation}
\begin{matrix}
\left|\Uparrow_{n};t\right\rangle=\frac{1}{\sqrt{2}}(-e^{-i\vartheta },1)^{T} & \\& \\
\left|\Downarrow_{n};t\right\rangle=\frac{1}{\sqrt{2}}(e^{-i\vartheta },1)^{T}
\end{matrix}
\end{equation}

As might be expected the phase, which can be obtained from (6) by making a re-gauge-definition appears to be independent of the spin characteristics of the electrons since the interaction term involves only electron charge and the gauge potential due to the solenoid. Then, each eigenstate in (7) picks up a topological AB phase factor $(e^{ie\Phi})$ at $t=\tau$ which does not depend on spin orientation of electrons.
\begin{equation}
\begin{matrix}
\left|\Uparrow_{n};t=0\right\rangle\to\left|\Uparrow_{n};t=\tau\right\rangle = e^{ie\Phi}\left|\Uparrow_{n};t=0\right\rangle & \\& \\
\left|\Downarrow_{n};t=0\right\rangle\to\left|\Downarrow_{n};t=\tau\right\rangle = e^{ie\Phi}\left|\Downarrow_{n};t=0\right\rangle
\end{matrix}
\end{equation}

Now we may consider an entangled two spin-1/2 particles in a singlet state to investigate the possible effects of the phase on entangled quantum system (a source creates an entangled pair of spin 1/2 particles as shown in Fig. 1). Furthermore, we can decompose the initial singlet state into the eigenstates of the interaction term in (6) to reveal the topological AB phase:

\begin{equation} \left|\psi (t=0)\right\rangle =\frac{1}{\sqrt{2}}\lbrace 
\left|\Uparrow_{n}\right\rangle_{L}\left|\Downarrow_{n}\right\rangle_{R}-\left|\Downarrow_{n}\right\rangle_{L}\left|\Uparrow_{n}\right\rangle_{R}\rbrace
\end{equation}

In this configuration one of the particles, e.g. moving through the rightward, interacts with the vector potential where the region there is no magnetic field while the other particle is not under the influence of any topological phase. Thus, just one of the subspaces of the Hilbert space is affected by the topological phase. Then, the state (9) takes the form at $t=\tau$,

\begin{figure*}
\includegraphics[width=\textwidth]{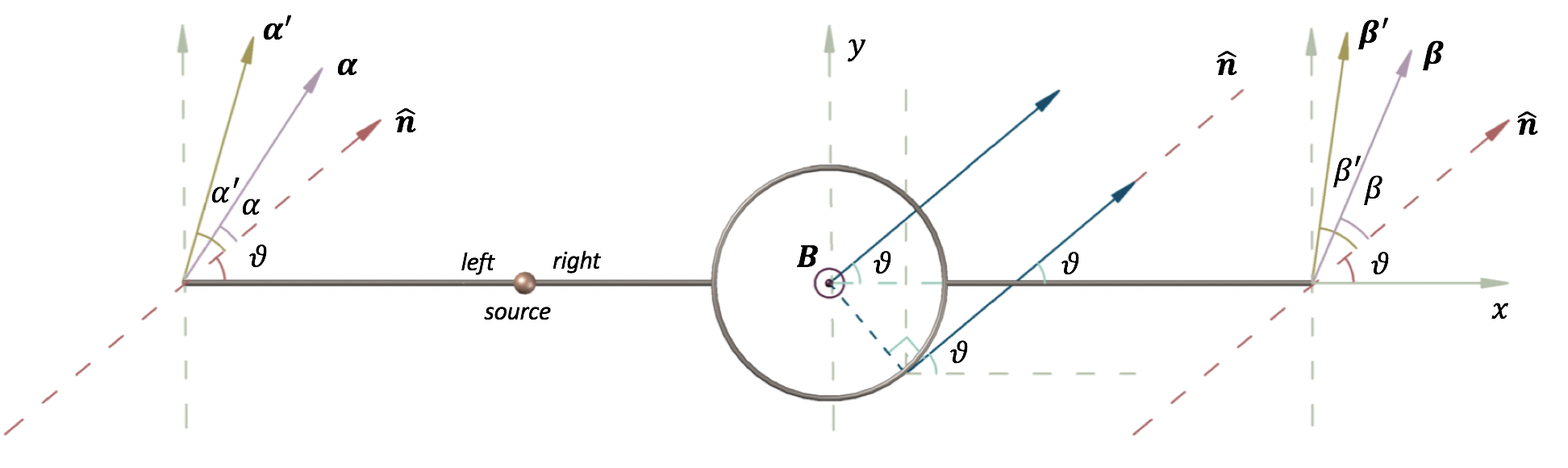}
\caption{\label{fig:wide2} A proposal for testing the possible effects of the AB phase on entangled quantum states.}
\end{figure*}

\begin{equation} \left|\psi (t=\tau)\right\rangle =\frac{1}{\sqrt{2}}e^{ie\Phi}\lbrace \left|\Uparrow_{n}\right\rangle_{L}\left|\Downarrow_{n}\right\rangle_{R}-\left|\Downarrow_{n}\right\rangle_{L}\left|\Uparrow_{n}\right\rangle_{R}\rbrace
\end{equation}

To examine the effects of the phase on entangled quantum states, one can measure the spin components of the left and right going particles simultaneously as in the well-known Bell experiments. But it is clear that $(e^{ie\Phi})$ in (10) is overall phase factor, then it can be neglected and the AB phase does not appear in quantum mechanical expectation value expressions in Bell type experiments. This result also illustrates that the AB phase is independent of the spin orientation of the electrons. On the other hand, while the role of geometric phase on entangled quantum state is examined with a similar setup in [25]\nocite{25}, the geometric Berry phase appears as a variable in their final expectation value expressions and correlation functions, since spin up and down components of the entangled state are acquiring different well known Berry phases ($\gamma^+ $ and $\gamma^-$) in an adiabatically rotating magnetic field. However in the AB problem, which is a special case of Berry phase, the spin components of the entangled state gain the same phase. Simply because the solenoidal magnetic field is always in one direction (namely, magnetic dipoles forming the infinitely thin solenoidal B field are polarized in two dimensions), besides the relevant interaction Hamiltonian does not have any spin dependence. For the AC effect discussed in the next section it will be seen that each spin component of the entangled state will get different phases and hence the phase appears as a variable in the final expectation values and correlation functions.

\section{The AC Efect in Entangled Quantum Systems}
\label{sec:sample3}

As is well known magnetic dipoles following the closed trajectory around the singular region created by the infinitely thin and long linear charge distribution, acquire the topological AC phase without the any effect of classical forces. To investigate this quantum mechanical phenomenon, it would be appropriate to start from the relativistic Lagrangian describing the dynamics of a neutral particle having solely magnetic dipole in an external electromagnetic field $F_{\mu \nu }$
\begin{equation}
L=\bar{\psi }(i\gamma ^{\mu }\partial _{\mu}-m+\frac{\mu }{2}\sigma ^{\mu \nu }F_{\mu \nu })\psi
\end{equation}

\noindent here $m$ and $\mu $ are the mass and magnetic dipole moment of neutral particle, and the corresponding equations of motion are,
\begin{equation}
(i\gamma ^{\mu }\partial _{\mu }-m+\frac{\mu 
}{2}\sigma ^{\mu \nu }F_{\mu \nu })\psi =0.
\end{equation}

When the problem is confined on a plane as in the previous section for AB case, the linear charge distribution $\lambda_{E}$ will reduce to a single electron, and the AC phase is formed as a result of unpolarized neutrons following a closed trajectory around an electric charge. Thus, the interaction term can be obtained by starting (3) as initial Hamiltonian and considering (4) and (5),

\begin{equation}
\Delta H_{AC}=-s\mu \boldsymbol{\alpha _{\perp}}\cdot\boldsymbol{\tilde{E}}.
\end{equation}

\noindent Here the $\boldsymbol{\tilde{E}}=\boldsymbol{E}\times \boldsymbol{\hat{z}}$ is redefined electric field vector with the components $(E_{y},-E_{x},0)$ in two-dimensions and $s$ originated from the two inequivalent representations of Dirac algebra in two dimensions which corresponds to spin up and down particles. Unlike equation (6) for the AB case, (13) exhibits a spin dependence for AC case. At this stage, the well known AC phase can be achieved as $(\chi _{AC}=s\mu \lambda _{E})$ with the elimination of interaction Hamiltonian by simply making a re-gauge-definition.  Hence, the instantaneous eigenstates of (13) along the $\boldsymbol{\tilde{n}}$ direction (in Fig. 2), 
\begin{equation}\begin{matrix}\left|\Uparrow_{\tilde{n}};t\right\rangle =\frac{1}{\sqrt{2}}(-e^{-i\vartheta },1)^{T}\\
\\
\left|\Downarrow_{\tilde{n}};t\right\rangle =\frac{1}{\sqrt{2}}(e^{-i\vartheta},1)^{T}
\end{matrix}\end{equation}

\noindent and the corresponding eigenvalues are $\lambda _{AC_{1,2}}=\pm \mu \tilde{E}$ with the definitions $\tilde{E}_{x}=\tilde{E}\cos{\vartheta} , \tilde{E}_{y}=\tilde{E} \sin{\vartheta}$ and $\tilde{E}^{2}=\tilde{E}_{x}^{2}+\tilde{E}_{y}^{2}$,  where $\vartheta $ is the angle between positive $x$ axis and arbitrary $\tilde{n}$ direction perpendicular to the electrical field vector ${\boldsymbol{E}}$ on $xy$ plane. Then each eigenstate gains a topological AC phase associated with their spin orientation after a complete cycle at $t=\tau $, 

\begin{equation}
\begin{matrix}
\left|\Uparrow_{\tilde{n}};t=0\right\rangle \to  \left|\Uparrow_{\tilde{n}};t=\tau\right\rangle = e^{-i\mu\lambda}\left|\Uparrow_{\tilde{n}};t=0\right\rangle & \\& \\
\left|\Downarrow_{\tilde{n}};t=0\right\rangle \to \left|\Downarrow_{\tilde{n}};t=\tau\right\rangle = e^{i\mu\lambda}\left|\Downarrow_{\tilde{n}};t=0\right\rangle
\end{matrix}
\end{equation}

The hitherto discussions provide a suitable basis for investigating how the topological phases appear in quantum entangled systems. For an EPR type problem, let us consider the initial singlet state (9) for a pair of magnetic dipoles (neutrons) moving along $\pm x$ directions; and denote the spin polarization axis to be $\boldsymbol{\hat{\tilde{n}}}$ (namely along the $\boldsymbol{\hat{\tilde{E}}}$, See Fig. 2). In this configuration the initial Bell singlet states in equation (9) now becomes at $t=\tau $

\begin{figure*}
\includegraphics[width=\textwidth]{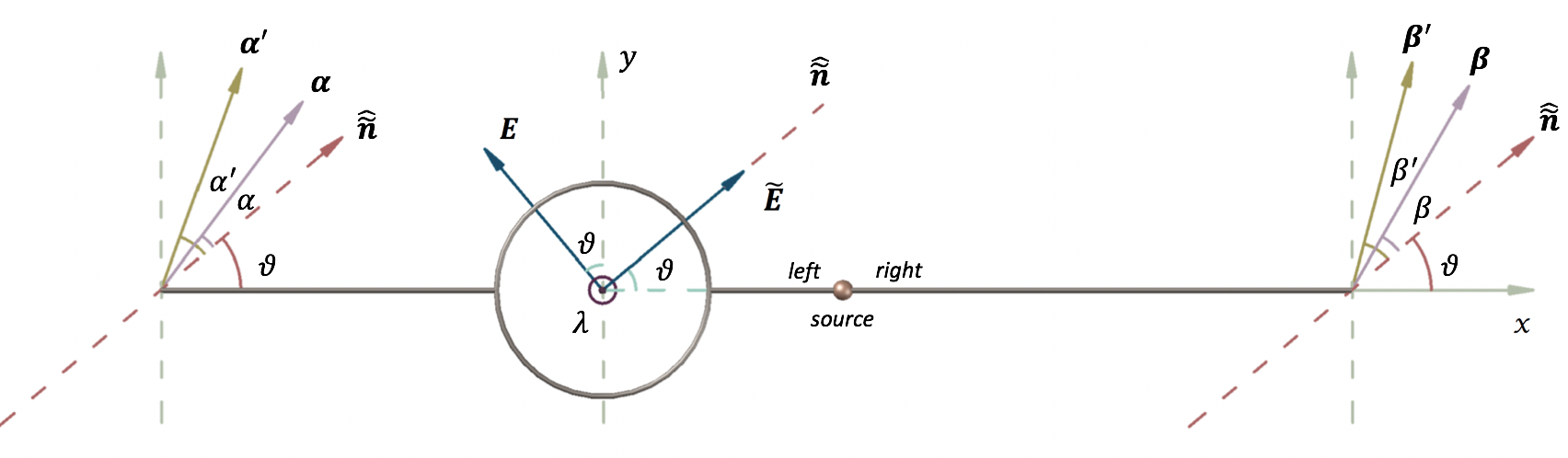}
\caption{AC effect for a pair of magnetic dipoles in an entangled quantum state. Unlike the conventional EPR type problem, here a new feature appears: The relation (17) becomes dependent on the topological phase when the left-going particles (neutrons), produced in entangled state from the source, are subjected to the AC effect. Thus correlation type non-local aspects of quantum mechanics can further be tested experimentally via the angles for which CHSH inequality is maximally violated by the quantum mechanical expectation values.}
\label{fig:boat1}

\end{figure*}

\begin{equation}\left|\psi (\tau )\right\rangle=\frac{1}{\sqrt{2}}\lbrace 
e^{-i\mu \lambda }\left|\Uparrow_{\tilde{n}}\right\rangle_{L}\left|\Downarrow_{\tilde{n}}\right\rangle_{R}-e^{i\mu \lambda }\left|\Downarrow_{\tilde{n}}\right\rangle_{L}\left|\Uparrow_{\tilde{n}}\right\rangle_{R}\rbrace\end{equation}

\noindent and it can be rewritten by neglecting overall phase factor,
\begin{equation} \left|\psi (\tau )\right\rangle =\frac{1}{\sqrt{2}}\lbrace 
\left|\Uparrow_{\tilde{n}}\right\rangle_{L}\left|\Downarrow_{\tilde{n}}\right\rangle_{R}-e^{2i\mu \lambda }\left|\Downarrow_{\tilde{n}}\right\rangle_{L}\left|\Uparrow_{\tilde{n}}\right\rangle_{R}\rbrace.\end{equation}

Just as in the well-known Bell experiments, one can investigate the result of the measurement of the spin components of the particles going left and right. For this purpose, let us define a projection operator along an arbitrary vector $\boldsymbol{\alpha}$ as shown in Fig. 2:
\begin{equation}P_{\pm }(\boldsymbol{\alpha})=\left|\pm\boldsymbol{\alpha}\right\rangle \left\langle \pm\boldsymbol{\alpha}\right|\end{equation}

\noindent $\left|+\boldsymbol{\alpha}\right\rangle$ and $\left|-\boldsymbol{\alpha}\right\rangle$ are spin up and down states in two-dimensions and the angle $\alpha $ is with the $\boldsymbol{\hat{\tilde{n}}}$-direction. Hence,

\begin{equation}
\begin{matrix}
\left|+\boldsymbol{\alpha}\right\rangle =  \frac{1}{\sqrt{2}}\lbrack \cos\frac{\alpha }{2}\left|\Uparrow_{\tilde{n}}\right\rangle+i\sin\frac{\alpha }{2}\left|\Downarrow_{\tilde{n}}\right\rangle\rbrack & \\& \\
\left|-\boldsymbol{\alpha}\right\rangle = \frac{1}{\sqrt{2}}\lbrack -\sin \frac{\alpha }{2}\left|\Uparrow_{\tilde{n}}\right\rangle+i\cos\frac{\alpha }{2}\left|\Downarrow_{\tilde{n}}\right\rangle\rbrack
\end{matrix}
\end{equation}

\noindent Similarly for the neutron moving to the right the spin measurement direction is chosen to be ${\boldsymbol{\hat{\beta}}}$, and the expressions for $\left|\pm\boldsymbol{\beta}\right\rangle$ are the same as (19) simply replacing $\alpha\leftrightarrow \beta.$ Then, the joint probabilities are found as 
\begin{eqnarray}
P(\alpha \Uparrow_{\tilde{n}},\beta \Uparrow_{\tilde{n}}) &=& P(\alpha \Downarrow_{\tilde{n}},\beta \Downarrow_{\tilde{n}})\nonumber\\
& = & \frac{1}{4}\lbrack 1-\cos\alpha\thinspace \cos\beta -\sin\alpha\thinspace \sin\beta\thinspace \cos(2\mu \lambda )\rbrack\nonumber\\
P(\alpha \Uparrow_{\tilde{n}},\beta \Downarrow_{\tilde{n}}) & = & P(\alpha \Downarrow_{\tilde{n}},\beta \Uparrow_{\tilde{n}})
\nonumber\\
& = & \frac{1}{4}\lbrack 1+\cos\alpha\thinspace \cos\beta +\sin\alpha\thinspace \sin\beta\thinspace \cos(2\mu \lambda )\rbrack
\end{eqnarray}
\noindent Introducing the observable
\begin{equation}A^{l}(\boldsymbol{\alpha})=P_{+}^{l}(\boldsymbol{\alpha} )-P_{-}^{l}(\boldsymbol{\alpha})\end{equation}

\noindent and similarly $B^{r}(\boldsymbol{\beta})$ for the angle {$\beta$}, we calculate the expectation value of joint measurement,
\begin{equation}
    E(\boldsymbol{\alpha},\boldsymbol{\beta}) = \left\langle \psi\left(\tau\right)\right|A^{l}(\boldsymbol{\alpha})\otimes B^{r}(\boldsymbol{\beta})\left|\psi\left(\tau\right)\right\rangle= -\cos\alpha\thinspace \cos\beta -\sin\alpha\thinspace \sin\beta\thinspace \cos(2\mu \lambda ).\thinspace\thinspace\thinspace\thinspace
\end{equation}

In order to test the non-local features one can make use of the CHSH inequality, which is a special case of BI, and the resulting S function is,

\begin{eqnarray} S(\alpha ,\beta ,\alpha ',\beta ')& = &\vert -\cos\alpha\thinspace \cos\alpha'-\sin\alpha\thinspace \sin\alpha'\thinspace \cos (2\mu \lambda )\nonumber\\
& + & \cos\alpha\thinspace \cos\beta +\sin\alpha\thinspace \sin\beta\thinspace \cos(2\mu \lambda )\vert\nonumber\\
& + &\vert -\cos\alpha'\thinspace \cos\beta'-\sin\alpha 
'\thinspace \sin\beta'\cos(2\mu \lambda )\nonumber\\
& - & \cos\beta\thinspace \cos\beta'- \sin\beta\thinspace \sin\beta' \thinspace \cos(2\mu \lambda )\vert
\end{eqnarray}

\noindent With the certain chosen angles S is maximally violated by the quantum mechanical expectation values,
\begin{equation}S(0,\frac{\pi }{4},\frac{3\pi }{4},\frac{\pi 
}{2},\delta _{AC})=\sqrt{2}+\sqrt{2}\vert \cos (2\mu \lambda )\vert \end{equation}

\noindent Equation (24) illustrates how the S function depends on the topological phase acting on the left-going particle produced in the entangled state in Fig. 2. The S function can be controlled by using topological phases hence correlation type non-local properties of quantum mechanics can be experimentally tested using neutrons with intrinsic magnetic dipole moments. On the other hand, in reference [25]\nocite{25} the role of the geometric phase on entangled quantum states in an adiabatically rotating magnetic field is examined in three dimensional space and their S function depends on the azimuthal angles and the geometric phase. Since the AC phase is a special case of the Berry phase, we observe that our result (24) is again reached when the azimuthal angles are fixed to zero in the corresponding results of the [25]\nocite{25}. Moreover, if the spin orientation of the moving particle had not have any role as in the AB case, the topological phase information would disappear between spin up and down states for the left and right going particles in the singlet state. Hence, (24) would take familiar value of $2\sqrt{2}$ as in the AB case. 
\begin{table}
\begin{center}\includegraphics[width=1\columnwidth,]{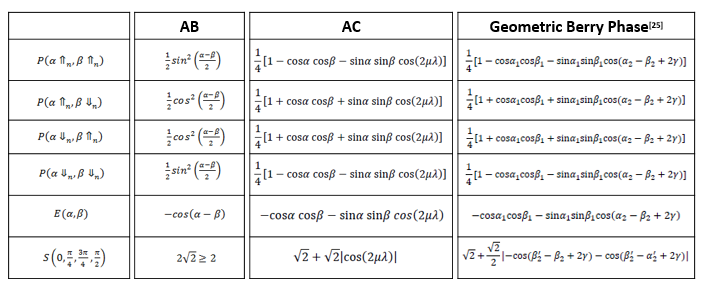}
\end{center}\caption{\label{fig:epsart} For entangled spin singlet state joint probabilities, expectation values of the joint measurement and the S functions of the CHSH inequality. When the azimuthal angles are set to zero in third column, the AC case coincide with the Berry phase case. For the AB case spin does not have any role and when $\mu\lambda$ is set to zero in second column (equivalent to letting $e\Phi=0$) one obtains AB case.}
\end{table}
But, it is clearly seen in equation (24) that AC phase appears in the S function, which is the natural consequence of the spin dependence of the AC Hamiltonian. The S functions for the AB, AC effects and Berry phase case for spin entangled quantum systems are summarized in the Table 1.

\begin{figure}
\begin{center}\includegraphics[width=0.7\columnwidth,]{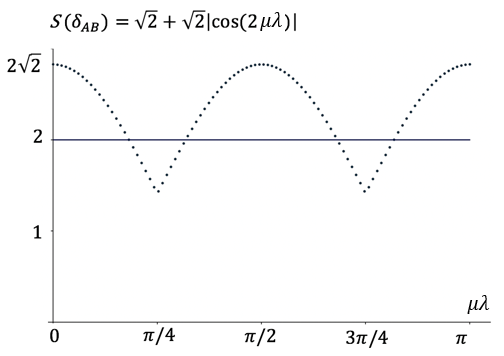}
\end{center}\caption{\label{fig:epsart} S function in CHSH inequality as a function of the AC phase}
\end{figure}

\section{Conclusion and Discussion}
\label{sec:sample3}

Topological AB and AC effects revolutionized our comprehension of the role of the electromagnetic potentials and very rich literature exists in this field. Here we analyzed these two effects in a two dimensional setting within a relativistic framework. Interaction Hamiltonians corresponding to each case are derived and the resulting geometric quantum phases turned out to have identical forms and coincide with the results already existing in the literature. Furthermore, the analysis is extended to entangled systems in the spirit of the work by Bertlmann et.al. [25]\nocite{25}, and CHSH inequality is investigated for both effects. Accordingly, it is understood that the effects of topological phases on the entangled quantum states may differ due to the dependence of the geometric phase on spin orientations of the moving particles. 

\begin{figure}
\begin{center}\includegraphics[width=1\columnwidth,]{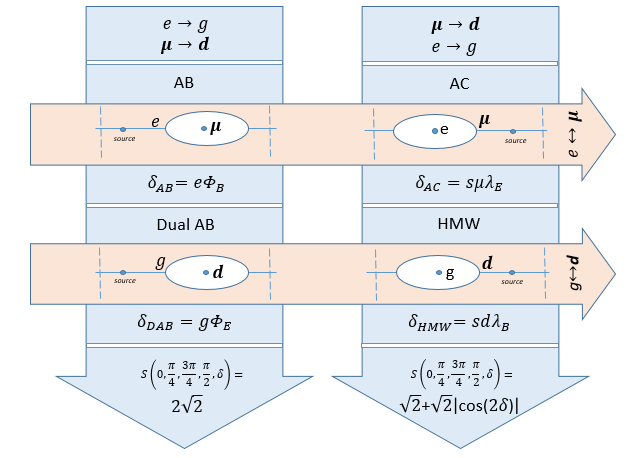}
\end{center}\caption{\label{fig:epsart} Summary of the geometric phases and S functions for entangled states in four type of the topological effects. Here e and g are electric and magnetic charges, $\boldsymbol{d}$ and $\boldsymbol{\mu}$ are electric and magnetic dipoles, $\Phi_E$ and $\Phi_B$ are electric and magnetic fluxes, $\lambda_E$ and $\lambda_B$ are electric and magnetic charge densities, and $s=\pm1$ indicates spin directions. We see that the AC phase and HMW phase which demonstrates a complete duality with AC simply by transforming $\boldsymbol{\mu}\rightarrow\boldsymbol{d}$ and $e\rightarrow g$ will exhibit the characteristic geometric topological phase dependence of the S function. However, the AB phase and the Dual AB phase (fourth topological phase which can be associated with a magnetic monopole around a string of electric dipoles demonstrates a complete duality with AB simply by transforming $e\rightarrow g$ and $\boldsymbol{\mu}\rightarrow\boldsymbol{d}$) do not appear to affect the S function.}
\end{figure}

\noindent Clearly, one can further apply a similar method to four possible vector type topological geometric phases and observe that only two of them AC and He-McKellar-Wilkens (HMW) effects \cite{46} exibit spin dependence and their correlation S functions for the entangled state do have dependence on the geometric phases. In contrary, for the other two cases, namely AB and its dual effect \cite{42}, have neither spin dependence and nor their S functions exhibit dependence on the geometric phase for the entangled state. Calculational details for two cases (HMW and Dual AB) and also the identity and the duality relationships among all of them will be presented in a future work.






\end{document}